\documentclass[letter]{aa} 
\usepackage{graphicx}
\usepackage{natbib}
\usepackage{epsfig}
\def\ut#1{\mathop{\vtop{\ialign{##\crcr
     $\hfil\displaystyle{#1}\hfil$\crcr\noalign
     {\kern1pt\nointerlineskip}\hbox{$\hfil\sim\hfil$}\crcr
     \noalign{\kern1pt}}}}}

\def\undersymbol#1#2{\mathop{\vtop{\ialign{##\crcr
     $\hfil\displaystyle{#2}\hfil$\crcr\noalign
     {\kern1pt\nointerlineskip}\hbox{$\hfil#1\hfil$}\crcr
     \noalign{\kern1pt}}}}}

\def\degr{^0}

\usepackage{graphicx}

\begin{document}

\title{Possible detection of the M31 rotation in WMAP data}
       \author{F. De Paolis\inst{1,2}, V.G. Gurzadyan\inst{3}, G. Ingrosso\inst{1,2}, Ph. Jetzer\inst{4},
       A.A. Nucita\inst{1,2}, A. Qadir\inst{5}, D. Vetrugno\inst{1,2}, A.L. Kashin\inst{3},
       H.G. Khachatryan\inst{3},  \and S. Mirzoyan\inst{3,4,6}}
              \institute{Dipartimento di Fisica, Universit\`a del Salento, via per Arnesano, 73100, Lecce, Italy  \and INFN, Sez. di Lecce, via per Arnesano, 73100, Lecce, Italy
              \and Yerevan Physics Institute and Yerevan State
University, Yerevan, Armenia
\and
Institut f\"ur Theoretische Physik, Universit\"at
Z\"urich, Winterthurerstrasse 190, 8057 Z\"urich, Switzerland
\and
Centre for Advanced Mathematics and Physics,
National University of Sciences and Technology, Rawalpindi,
Pakistan
\and
Dipartimento di Fisica ``E.R. Caianiello'',
Universit\`a degli Studi di Salerno, Fisciano, Italy
}

   \offprints{F. De Paolis, \email{francesco.depaolis@le.infn.it}}
   \date{Submitted: XXX; Accepted: XXX}


  \abstract
   {
Data on the cosmic microwave background (CMB) radiation by the
Wilkinson Microwave Anisotropy Probe (WMAP) had a profound impact
on the understanding of a variety of physical processes in the
early phases of the Universe and on the estimation of the
cosmological parameters. 
}
   {Here, the 7-year WMAP data
are used to trace the disk and the halo of the nearby giant spiral
galaxy M31. }
   {We analyzed the temperature
excess in three WMAP bands (W, V, and Q) by dividing the region of
the sky around M31 into several concentric circular areas. We
studied the robustness of the detected temperature excess  by considering
500 random control fields in the real WMAP maps and simulating 500
sky maps from the best-fitted cosmological parameters. By
comparing the obtained temperature contrast profiles with the real
ones towards the M31 galaxy, we  find that the temperature
asymmetry in the M31 disk is fairly robust, while the effect in
the halo is weaker. }
   {An
asymmetry in the mean microwave temperature in the M31 disk along
the direction of the M31 rotation is observed with a temperature
contrast up to $\simeq 130~\mu$K/pixel. We also find a temperature
asymmetry in the M31 halo, which is much weaker than for the disk,
up to a galactocentric distance of about $10\degr$ ($\simeq 120$
kpc) with a peak temperature contrast of about $40~\mu$K/pixel. }
   { Although the confidence level of the signal is not high, if estimated purely
   statistically, which could
be expected due to the weakness of the effect, the geometrical
structure of the temperature asymmetry points towards a definite
effect modulated by the rotation of the M31 halo. This result
might open a new way to probe these relatively less studied
galactic objects using high-accuracy CMB measurements, such as
those with the Planck satellite or planned balloon-based
experiments, which could prove or disprove our conclusions. }


   \authorrunning{De Paolis et al.}
   \titlerunning{Possible detection of the M31 rotation in WMAP data}
   \maketitle
%

\section{Introduction}
Galaxy rotation, in particular for the Andromeda galaxy (M31) has
been  well studied  especially in the optical, IR, and radio
bands, and it gives important information on the mass distribution
not only in galactic disks but also in their halos \citep{binney}.
On the other hand, since they are not directly observable, but
their presence is deduced from their effect on galactic dynamics,
galactic halos are relatively less studied structures of galaxies.
Various populations, such as globular clusters, RR Lyrae,
subdwarfs, and other types of stars, have been used to trace the
halo of the Galaxy,  its vertical structure, and its rotation
speed \citep{Kinman}. Nevertheless, there are still many
ambiguities not only in the main halo constituents, but also in
the basic properties such as, in particular, in rotation.

The degree to which galactic halos rotate with respect to the
disks is difficult to investigate; actually, as stated in the most
recent study of M31 \citep{courteau}, testing for the rotation of
M31's halo is still beyond our reach. Naturally, the importance of
understanding the galactic halos is closely related to the nature
and distribution of the dark matter, which is relevant for the
formation and dynamics of galaxies. In this respect, the
methodology adopted in the present paper of using WMAP data to
probe both the disk and the halo of M31, even if with the
limitation of the presently available data, may suggest a novel
way of approaching this problem.


\section{The 7-year WMAP analysis}
\begin{figure}[h]
\resizebox{\hsize}{!}{\includegraphics[angle=0,width=62mm,width=0.9\textwidth]{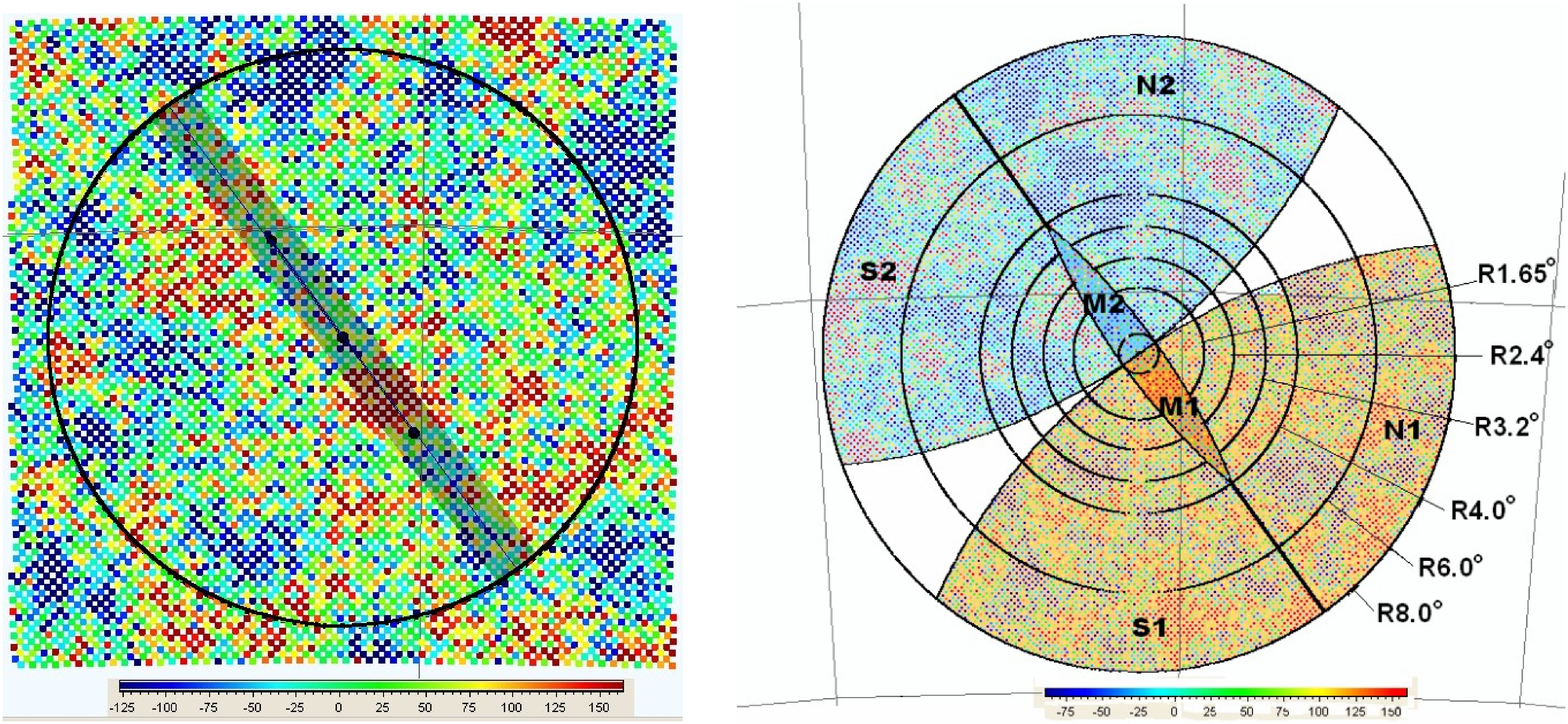}}
\caption{In the left panel, the WMAP W-band towards the M31
galaxy. The $8.5\degr\times 8.5\degr$ sky field centered at
$(121.17\degr, -21.57\degr)$ with the marked $4\degr$ circular
region. The oblique strip indicates the M31 disk, and the analysis
in the halo region of M31 galaxy is extended far beyond the region
indicated in the figure. The detailed geometry (up to $8\degr$)
used in the analysis is shown in the right panel.} \label{fig1}
\end{figure}
In our analysis we use the seven-year WMAP data \citep{J} in the
three bands W (94 GHz),  V (61 GHz), and Q (41 GHz). Using three
WMAP bands is important in revealing the possible contribution of
the Galactic foregrounds since dust, free-free, and synchrotron
emission contributes differently in each band. Here we remind the
reader that the band least contaminated by the synchrotron
radiation of the Galaxy is the W-band, which also has the highest
angular resolution. The CMB map's general structure in the W-band
in the region of M31, with the marked $4\degr$ radius circle
(although our analysis extends farther out), is shown in Fig.
\ref{fig1} (left panel). In our analysis, we also used the maps
provided by the WMAP Collaboration with the Galactic disk
contribution modeled and removed \citep{gold}. It is always
specified in the text when we considered these data. To reveal the
different contributions by the M31 disk and halo, the region of
the sky around the M31 galaxy was divided into several concentric
circular areas as shown in Fig. \ref{fig1} (right panel). In the
optical band the total extent of the M31 galaxy along the major
axis is slightly more than  about $3\degr$ and along the minor
axis is about  $1\degr$. Radio observations have shown that the
M31 HI disk is more extended with respect to the stellar disk
\citep{Chemin,Corbelli}, with a major axis sizes of about
$5.6\degr$ and a minor axis size of about $1.2\degr$. In this
paper, the adopted M1 and M2 disk regions (Fig. \ref{fig1}, right
panel) have major axis size of $8\degr$ and minor axis size of $1
\degr$; this allows us to retain the warped part of the M31 disk
in the M1 and M2 regions. Moreover, we have checked that it is
irrelevant, as far as our analysis is concerned, to extend the M31
minor axis to $1.2\degr$. The mean temperature excess per pixel
$T_m$, in $\mu$K/pixel, in each region was obtained in each band
and is shown in Table 1 with the $1\sigma$ error \footnote{The
standard error of the mean is calculated as the standard deviation
of the excess temperature distribution divided by the square root
of the pixel number in each region. We have verified that, within
the errors, the sigma values calculated in that way are consistent
with those evaluated by using the covariance matrix obtained by a
best-fitting procedure with a Gaussian to the same distribution.},
along with the number of pixels in each area. For convenience,
Table 1 gives the temperature excess in each M31 region up to
$8\degr$, even if our analysis was extended to the region around
the M31 disk with concentric annuli with radii up to $20\degr$.
\begin{table*}[tbp]
\centering \caption{Temperature excess in the M31 regions for the
non foreground-reduced WMAP maps.}
\begin{tabular}{clrccc}
\hline \hline
 R, deg, kpc & Region       & N, pix  &   W, $T_m\pm \sigma$ &  V, $T_m\pm\sigma$ &   Q, $T_m\pm\sigma$ \\
\hline
 ~~~ 1.65, 21.4      & N1 + M1 + S1 &   324   &    $63.1\pm 5.6$ &   $67.2\pm 5.4$ &     $90.0\pm 4.2$ \\
           & N2 + M2 + S2 &   321   &    $20.3\pm 4.7$ &  $ 17.3\pm 4.3$ &     $ 37.0\pm 3.3$ \\
           & N1 + S1      &   205   &45.5$~\pm$  5.7 &   38.0$~\pm $ 5.3&   64.1$~\pm$  4.0 \\
           & N2 + S2      &   205   & 33.8$~\pm$  5.9 &   34.3$~\pm$  5.3 &   41.8$~\pm$  4.1\\
           & M1           &   119   &   $121.4\pm 19.6$ &  $117.6\pm 10.0$ &  $134.3\pm 7.4$ \\
           & M2           &   116   &    $-7.7\pm 7.4$ & $-12.7\pm 6.8$ & $28.4\pm 5.5$ \\
\hline
 ~~~ 2.40, 31.1    & N1 + M1 + S1 &   670    &   $43.7\pm 3.6$ &  $43.5\pm 3.4$ &  $66.0\pm 2.8$ \\
           & N2 + M2 + S2 &   664    &   $21.0\pm 3.6$ &   $19.1\pm 3.3$ &  $35.6\pm 2.7$ \\
           & N1 + S1      &   506    &  41.0$~\pm$  3.5 &   32.2$~\pm$ 3.1 &  55.6$~\pm$  2.6 \\
           & N2 + S2      &   504    &  24.5$~\pm$ 4.3 &   23.6$~\pm$  3.9 &  34.8$~\pm$  3.3 \\
           & M1           &   164    &   $73.0\pm 9.4$ &  $78.3\pm 9.3$ & $98.2\pm 7.4$ \\
           & M2           &   160    &    $11.2\pm 6.3$ &  $5.8\pm 6.4$ &  $38.1\pm 4.8$ \\
\hline
 ~~~ 3.20, 41.5      & N1 + M1 + S1 &  1176    &   $36.5\pm 2.7$ &  $38.9\pm 2.4$ & $59.6\pm 2.1$ \\
           & N2 + M2 + S2 &  1166    &   $16.0\pm 2.7 $ &  $11.3\pm 2.5$ & $30.3\pm 2.2$ \\
           & N1 + S1      &   980    &  35.3$~\pm$  2.7&    34.0$~\pm$  2.3 &  53.9$~\pm$  2.1 \\
           & N2 + S2      &   976    &  16.4$~\pm$  3.0 &   11.5$~\pm$  2.8 &  27.7$~\pm$  2.4 \\
           & M1           &   196    &   $63.8\pm 8.4$ &  $63.2\pm 8.3$ &  $88.1\pm 6.6$ \\
           & M2           &   190    &   $12.5\pm 5.8$ &  $10.3\pm 5.8$ & $44.0\pm 4.4$ \\
\hline
~~~ 4.00, 51.9      & N1 + M1 + S1 &  1818    &   $37.4\pm 2.2$ &   $39.6\pm 2.0$ &  $56.7\pm 1.7$ \\
           & N2 + M2 + S2 &  1808    &    $1.7\pm 2.3$ & $-2.4\pm 2.1$ & $16.9\pm 1.9$ \\
           & N1 + S1      &  1610    &  36.8$~\pm$  2.2 & 36.9$~\pm$  1.9 &    53.3$~\pm$  1.7 \\
           & N2 + S2      &  1609    & $-$ 0.4$\pm$ 2.5 &  $-$ 3.8$~\pm$  2.2 &   13.5$~\pm$  1.9 \\
           & M1           &   208    &   $64.7.0\pm8.1$ & $60.8\pm 7.9$ &  $83.6\pm 6.4$ \\
           & M2           &   200    &   $12.5\pm 5.6$ & $9.4\pm 5.6$ &  $43.6\pm 4.3$ \\
\hline
~~~ 6.00, 77.8      & N1 + S1      &  3748    &  29.7$~\pm$  1.5 &    27.0$~\pm$  1.4 &   44.0$~\pm$ 1.2 \\
                    & N2 + S2      &  3749    &  11.3$~\pm$  1.7 &   7.1$~\pm$  1.5 &   25.8$~\pm$  1.3 \\
\hline
~~~ 8.00, 103.8     & N1 + S1      &  6606    &  34.3$~\pm$  1.2 &   34.7$~\pm$  1.1 &  51.5$~\pm$  4.0 \\
                    & N2 + S2      &  6600    &  19.2$~\pm$  1.3 &   15.0$~\pm$  1.2 &  38.7$~\pm$  1.0 \\
\hline
\end{tabular}
\tablefoot{The radius of the considered annulus is given in
degrees and in kpc in the first column; the value of 744 kpc
\citep{vilardell} is adopted for the distance to M31. The
considered regions as in Fig. 1 (right panel). The numbers of
pixels in each region are given. The last three columns show the
CMB mean temperatures per pixel of each region in $\mu$K in the W,
V, and Q bands, respectively, with the corresponding $1\sigma$
errors (see text for details).}
\end{table*}

\section{Results for the M31 disk}

For the M31 disk, our analysis shows that each M1 region is always
hotter than the corresponding M2 region, as can be seen from Table
1. Indeed we find a temperature excess contrast (i.e. the
difference between the temperature excesses per pixel) between the
M1 and M2 regions in all three WMAP bands that turns out to be
about $129\pm 21~\mu$K/pixel within 21.4 kpc (in the W band) and
then slightly decreases (but remains as large as about $41\pm
10~\mu$K/pixel at about 50 kpc). This effect seems to come from
the rotation induced Doppler shift of the gas and dust emission
from the M31 disk - indeed, the hotter (M1) region corresponds to
the side of the M31 disk that rotates towards us. \footnote{A
detailed study of the frequency dependent temperature asymmetry in
the CMB arising from different distributions of gas and dust in
the M31 disk is left to a forthcoming paper. In any case, although
some inhomogeneity in the disk structure is not excluded, there is
no reason to assume that it is the sole cause.} If one compares
what WMAP data show towards the M1 and M2 regions with the maps of
the M31 thick HI disk obtained at 21 cm \citep{Chemin,Corbelli}
one sees a remarkable superposition of the hot (M1) and cold (M2)
regions in both observations.
\begin{figure}[h]
\resizebox{\hsize}{!}{\includegraphics[angle=0,width=54mm,width=0.85\textwidth]{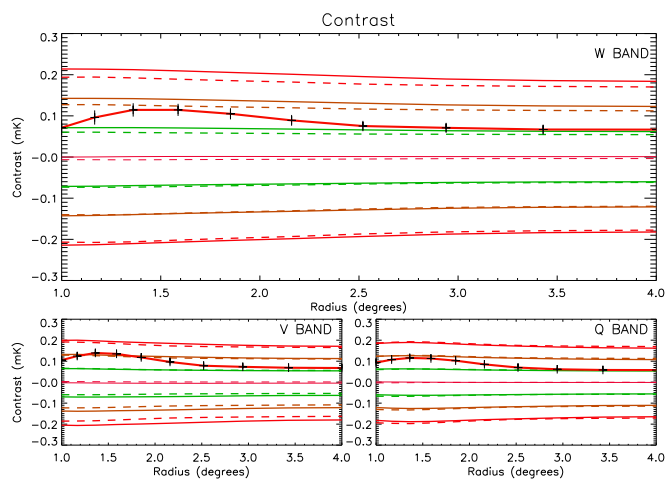}}
\caption{The 1$\sigma$ (green lines), 2$\sigma$ (brown lines), and
3$\sigma$ (red lines) excess temperature contrast (in mk/pixel)
curves (in the W, V, and Q bands) along with the mean profile
(pink line close to zero) for 500 random control fields
(continuous lines) and 500 simulated CMB sky maps (dashed lines).
In red, the observed temperature contrast profile in the M31 disk
(with $1\sigma$ errors) is given. The WMAP maps with the Galactic
disk contribution modeled and removed (foreground-reduced maps)
are used here.} \label{f2}
\end{figure}
Even if the temperature asymmetry between the M1 and the M2
regions looks significant, we have to check whether it might be
due to a random fluctuation of the CMB signal. It is indeed well
known that the CMB sky map has a ``patchy'' structure
characterized by the presence of many hot and cold spots with
temperature excesses up to several tents of $\mu$K/pixel on
angular degree scales. We therefore considered (Fig. \ref{f2} and
also the online material) 500 control fields and 500 simulated sky
maps  (from the best-fitted cosmological parameters as provided in
the WMAP web site) \footnote{CMB maps were simulated by assuming
$\Delta T(\hat{n}) = \Delta T_{CMB}(\hat{n})\otimes B(\hat{n}) +
N(\hat{n})$, where $\Delta T_{CMB}$ is a realization of the
Gaussian CMB field, $N(\hat{n})$ is the pixel noise and
$B(\hat{n})$ is the proper beam of the experiment. We made 500
realizations of the CMB sky using {\tt synfast} routine of HEALPix
with the best-fit power spectrum as given by the WMAP
Collaboration. The maps are then convolved with WMAP beams for W,
V, and Q bands, respectively. Noise realizations (simulated with
$\sigma_0 = 6.549$ mK, $\sigma_0 =3.137$ mK,  and $\sigma_0
=2.197$ mK for W, V, and Q-bands, respectively) are added to the
beam convolved maps in the end.} - and evaluated the temperature
contrast with the same geometry as was used towards M31. We also
give (red curve) the M31 temperature contrast profile in the M31
disk. Due to our chosen geometry, each curve is given up to
$4\degr$. As one can see, the contrast temperature profile for the
M31 disk is always a nicely smooth curve that is close to the
$2\sigma$ curve in the intermediate region of about $1.5-2\degr$.
Both the control field and the CMB simulation analyses show that
there is a probability of less than about $4\%$ that the
temperature asymmetry revealed comes from a random fluctuation of
the CMB signal. Actually, if one takes the direction of rotation
of the M31 disk into account, such a probability reduces (by using
the theorem of the composite probability) by a factor of two.
Finally, we mention that we have found that the temperature excess
contrast of the two M31 disk regions obtained by dividing the
M1+M2 region with respect to the north-west/south-east symmetry
axis (the M31 disk axis) turns out to be $0.008\pm 0.012$ mK,
which seems to further confirm that the temperature contrast
between the M1 and M2 regions is not due to a random fluctuation
in the CMB signal.

\section{Results for the M31 halo}
The next step was to enlarge our analysis to the region around the
M31 disk by considering concentric circular regions of increasing
galactocentric radii (see also Fig. 1, right panel). We estimated
the difference of the temperature excess in the region N1+S1 in
the three WMAP bands with respect to that in the region N2+S2. A
temperature contrast between the region N1+S1 with respect to
N2+S2 shows up (see Table 1), and the N1+S1 region turns out to
always be hotter than the N2+S2 region. The detected effect
resembles the one towards the M31 disk, although with less
temperature asymmetry. In all three bands, the maximum temperature
contrast reaches a maximum at a galactocentric distance of about
$4\degr$ and then decreases slightly. It is apparent from the size
of the considered regions that a contamination of the M31 disk in
the regions N1, N2, S1, and S2 can be completely excluded, and
also the Galactic plane emission cannot account for the observed
temperature asymmetry  since it eventually would make a larger
contribution towards the upper regions of M31 (while the opposite
is observed in the data). As for the M31 disk, the temperature
asymmetry in the M31 halo is indicative of a Doppler shift
modulated effect possibly induced by the rotation of the M31 halo.

\begin{figure}[h]
\resizebox{\hsize}{!}{\includegraphics[angle=0,width=54mm,width=0.85\textwidth]{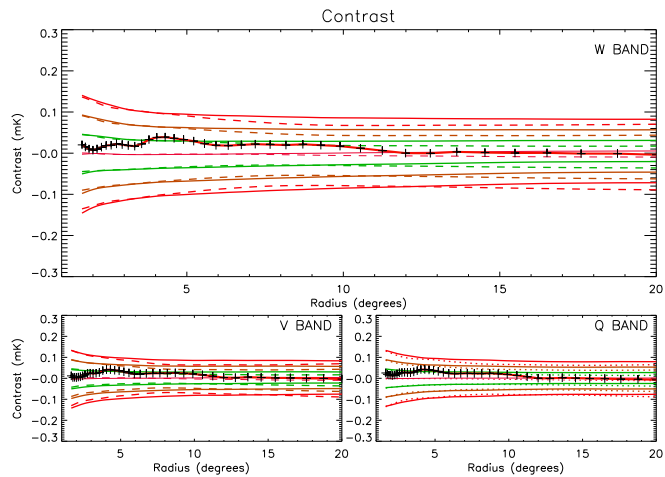}}
\caption{The same as Fig. 2 but for the M31 halo (temperature
contrast of the N1+S1 with respect to the N2+S2 regions in
mK/pixel - red line) for 500 random control fields (continuous
lines) and 500 simulated sky maps (dashed lines). Here the WMAP
maps with the Galactic disk contribution modeled and removed
(foreground-reduced maps) are used. } \label{f3}
\end{figure}

Also in this case we need to check the robustness of our results;
that is, we have to estimate the probability that the temperature
asymmetry in the M31 halo is due to a random fluctuation of the
CMB signal. In Fig. \ref{f3} (see also the online material) we
have considered 500 control fields and 500 simulated sky maps
(from the best-fitted cosmological parameters as provided on the
WMAP website). As one can see, in all three bands, the contrast
temperature profile of the M31 halo is close to the $1\sigma$
curve up to about $10\degr$ and goes slightly beyond it at about
$50-60$ kpc where the halo effect is maximum. This means that
there is less than $30\%$ probability that the temperature
contrast we see towards the M31 halo is due to a random
fluctuation of the CMB signal.
We also point out that we have verified that the temperature
asymmetry towards the M31 halo vanishes if the adopted geometry is
rigidly rotated of an angle larger than about $10\degr$ with
respect to the M31 major axis, thus giving a further indication
that the halo temperature contrast effect might be genuine and not
simply a random fluctuation of the CMB.

We also point out that the use of three WMAP bands is useful for
revealing the role of the contribution to the Galactic foregrounds
since each emission mechanism contributes differently in each
band. That the temperature contrast seems present in all three
bands and is more or less the same in each band up to about
$10\degr-11\degr$ indicates that the foregrounds are far weaker
than the effect. This size corresponds to the typical size
inferred for the dark matter halos around massive galaxies and
might open the possibility of a new way of studying these systems,
galactic disks, and halos, by using the microwave band. In any
case, a careful analysis of the Planck data that will be released
shortly should allow either proving or disproving our results.

\section{Discussion and conclusions}
We have shown that a temperature asymmetry  in all WMAP bands may
exist both in the M31 disk and halo in the direction of the M31
spin. For the M31 disk, the effect is fairly clear, and there is a
probability below about $2\%$ that it is a random fluctuation of
the CMB signal. If real, the detected temperature excess asymmetry
should be due to the foreground emission of the M31 disk modulated
by the Doppler shift of the disk spin. That the present study is
really timely is strengthened  by considering that the M31 galaxy
is already detected by the Planck observatory \citep{planckm31}
\footnote{However, there is no mention of any temperature
asymmetry in the M31 disk in that paper.}, whereas it did not
appear in the WMAP list. These are all reasons to expect that the
particular effect we discuss here can be studied more accurately
with Planck data.

As for the M31 halo, we have shown that, although less evident
than for the M31 disk, there is some evidence of a temperature
asymmetry between the N1+S1 and the N2+S2 regions that resembles
that of a Doppler shift effect induced by the M31 halo rotation.
We have shown in the previous section that there is less than
about $30\%$ probability that the detected temperature asymmetry
at a galactocentric distance $\sim 50$ kpc comes from a random
fluctuation of the CMB signal. \footnote{We  also mention that the
number and the temperature profile of radio sources in CMB maps
\citep{gurzadyan} excludes their significant contribution in the
effect under study.}

If one assumes that this temperature asymmetry in the M31 halo
relies in the M31 itself and is related to the M31 halo rotation,
one could speculate about the origin of this effect. In general,
four possibilities may be considered: (a) free-free emission, (b)
synchrotron emission, (c) Sunyaev-Zel'dovich (SZ) effect, and (d)
cold gas clouds populating the M31 halo \footnote{We  also
considered the possible influence of the observed high-velocity
clouds, either in the M31 or in our galaxy halos
\citep{Westmeier,Hulsbosch,Morras}, by removing the pixels in the
direction of each cloud from the analysis. The results obtained do
not change with respect to those presented here, as expected when
also considering the relatively low number of pixels involved.
Also the proposed ejecta by the past interaction of M33 and M31
galaxies \citep{Bekki} cannot play any role in our analysis since
it would at most have made hotter some pixels in the S2 region
(where the M31-M33 bridge is located), which is instead colder
than the S1 one.}. To work, the first three effects, assume the
presence of a rather hot plasma in the halo of M31. Although this
hot plasma has not been detected yet, one can assume that a
certain amount of this plasma  can populate the M31 halo (spiral
galaxies are believed to have much less hot gas than ellipticals)
and may rotate with a certain speed. Free-free emission arises
from electron-ion scattering while synchrotron emission comes
mostly from the acceleration of cosmic-ray electrons in magnetic
fields. Both effects give rise to a thermal emission with a rather
steep dependence on the frequency \citep{bennett} that therefore
should give a rather different temperature contrast in the three
WMAP bands. The absence of this effect indicates that the
contribution from possibilities (a) and (b) should be negligible.
And for (c), even for typical galaxy clusters with diffuse gas
much hotter than that possibly expected in the M31 halo, the
rotational scattering effect would produce a temperature asymmetry
of at most a few $\mu$K/pixel, depending on the rotational
velocity and the inclination angle of the rotation axis
\citep{cooray}. Actually, a possible temperature asymmetry in the
CMB data towards the M31 halo as a consequence of the existence of
a population of cold gas clouds in its halo was predicted in
\citep{paperdijqr95} - possibility (d). Indeed, if the halo of the
M31 galaxy contains cold gas clouds, we expect them to rotate like
the M31 disk (even if, perhaps, more slowly), and thus there
should be a Doppler shift inducing a temperature anisotropy
$\Delta T$ between one side of the M31 halo and the other with
respect to the rotation axis perpendicular to the disk. In the
case of optically thin halo clouds, the Doppler induced
temperature anisotropy would be ${\Delta T}/{T_r} \simeq
2{v}S~\bar \tau /c~$, where  $v$ is the M31 rotation speed, $\bar
\tau$ the averaged cloud optical depth over the frequency range
($\nu_1 \leq \nu \leq \nu_2$) of a certain detection band, and $S$
the cloud filling factor, i.e. the ratio of filled (by clouds) to
total projected surface in a given field of view. We emphasize
that the fact that the temperature contrast in Fig. \ref{f3} looks
approximately the same in each band makes a point towards either
possibility (d) or a random fluctuation of the CMB sky (but with a
probability, if estimated purely statistically, of less of $30\%$
for the last possibility).

The wealth of data especially in the last decade shows that there
is good evidence for the presence in the halos of spiral galaxies
of gas in all gaseous phases: neutral, warm atomic, and hot X-ray
emitting gas  \citep{bregman}. Atomic gas (often identified as
HVCs) is observed in the radio band (particularly at 21 cm) and
through absorption lines towards field stars and quasars. The hot
gas may be detected in X-rays, while searches for cold gas clouds
in galactic halos are more problematic as are searches for them by
the presence of a gamma-ray halo \citep{dixon,depaolis99}, stellar
scintillations \citep{moniez03,habibi}, obscuration events towards
the LMC \citep{drakecook}, ortho-$H_2D^+$ line at 372 GHz
\citep{ceccarelli}, and extreme scattering events in quasar
radio-flux variations \citep{walker} have given no clear
indication of their presence.

In conclusion, we showed that our analysis based on seven-year
WMAP data suggests there is a temperature excess asymmetry in the
M31 disk is likely due to the M31 foreground emission modulated by
the Doppler shift induced by the M31 spin. We find that there is
less than $\simeq 2\%$ probability that the signal up to about 20
kpc  comes from a random fluctuation in the CMB signal. For the
M31 halo, we also find a temperature excess asymmetry between the
N1+S1 and the N2+S2 regions along the expected spin direction,
suggestive of a rotation induced Doppler shift. The effect in the
M31 halo is far weaker than for the disk, as obviously expected,
and more precise data are necessary before drawing any firm
conclusion. In all cases, this research may open a new window into
the study of galactic disks and especially the rotation of
galactic halos by using the Planck satellite or planned
balloon-based experiments.

\acknowledgements{We acknowledge the use of the Legacy Archive for
Microwave Background Data Analysis (LAMBDA). Support for LAMBDA is
provided by the NASA Office of Space Science.  Some of the results
in this paper were derived using the HEALPix
\citep{gorski/etal:2005} package. PJ acknowledges support from the
Swiss National Science Foundation. An anonymous referee is also
acknowledged.}


\Online

\begin{figure}
 \centering
\resizebox{\hsize}{!}{\includegraphics[angle=0,width=54mm,width=0.85\textwidth]{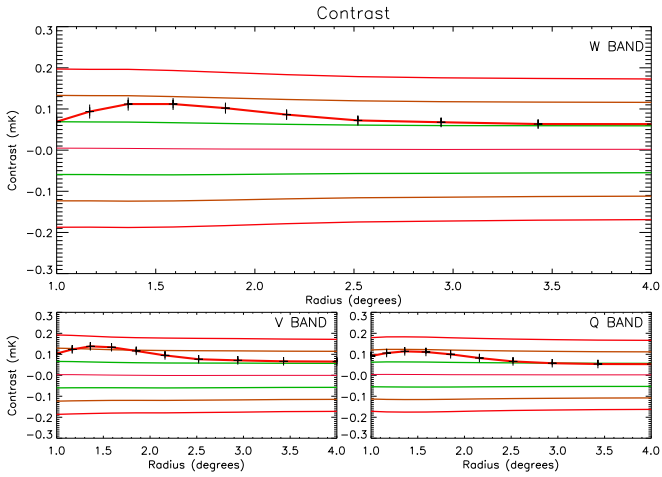}}
\caption{In the W, V, and Q bands, the 1$\sigma$ (green lines),
2$\sigma$ (brown lines), and 3$\sigma$ (red lines) excess
temperature contrast (in mk/pixel) curves, along with the mean
profile (pink line close) for 500 random control fields. The
observed temperature contrast profile in the M31 disk (with
$1\sigma$ errors) is given in red. The non foreground-reduced WMAP
maps are used here.} \label{appfig1}
\end{figure}

\begin{figure}
\centering
\resizebox{\hsize}{!}{\includegraphics[angle=0,width=54mm,width=0.85\textwidth]{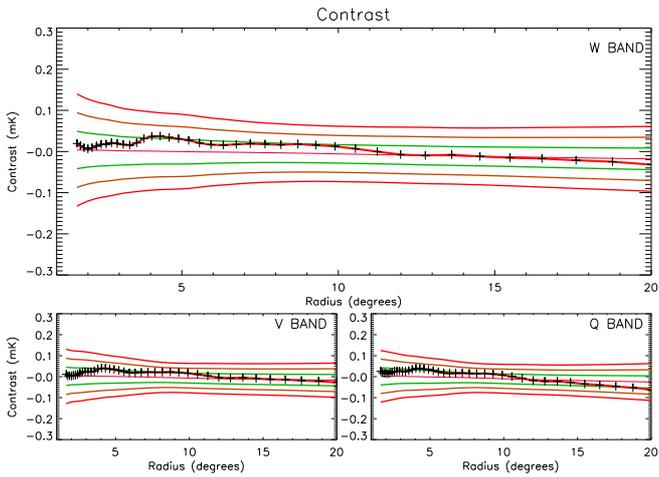}}
\caption{As above (in the W, V, and Q bands) the 1$\sigma$ (green
lines), 2$\sigma$ (brown lines), and 3$\sigma$ (red lines) excess
temperature contrast (in mk/pixel) curves along with the mean
profile (pink line close) for 500 random control fields. The real
temperature contrast profile in the M31 halo up to $20\degr$ (with
$1\sigma$ errors) is given in red.} \label{appfig2}
\end{figure}

\end{document}